\documentclass[10pt]{iopart}

\usepackage{iopams} 
\usepackage{graphicx}
\usepackage{float}
\usepackage[caption=false]{subfig}
\usepackage[export]{adjustbox}
\usepackage{placeins}
\usepackage{hyperref}
\usepackage{siunitx}
\usepackage{amsmath}
\usepackage[dvipsnames]{xcolor}
\usepackage[columnwise]{lineno}
\usepackage[T1]{fontenc}
\usepackage{selinput}

\definecolor{correction}{rgb}{1.0, 0.03, 0.0}
\definecolor{discuss}{rgb}{1.0, 0.28, 1.0} 

\begin{document}

\title[]{Effect of iron thicknesses on spin transport in a Fe/Au bilayer system}
\author{J. Briones$^1$, M. Weber$^1$, B. Stadtmüller$^{1,2}$, H. C. Schneider$^1$  and B. Rethfeld$^1$}

\address{$1$ Department of Physics and OPTIMAS Research Center, RPTU Kaiserslautern-Landau, Gottlieb-Daimler-Str. 76, 67663 Kaiserslautern, Germany}
\address{$2$ Institute of Physics, Johannes Gutenberg University Mainz, 55128 Mainz, Germany}

\ead{briones@physik.uni-kl.de}
\vspace{10pt}

\begin{abstract}
This paper is concerned with a theoretical analysis of the behavior of optically excited spin currents in bilayer and multilayer systems of ferromagnetic and normal metals. As the propagation, control and manipulation of the spin currents created in ferromagnets by femtosecond optical pulses is of particular interest, we examine the influence of different thicknesses of the constituent layers for the case of electrons excited several electronvolts above the Fermi level. Using a Monte-Carlo simulation framework for such highly excited electrons, we first examine the spatio-temporal characteristics of the spin current density driven in a Fe layer, where the absorption profile of the light pulses plays an important role. Further, we examine how the combination of light absorption profiles, spin-dependent transmission probabilities, and iron layer thicknesses affect spin current density in a Fe/Au bilayer system. For high-energy electrons studied here, the interface and secondary electron generation have a small influence on spin transport in the bilayer system. However, we find that spin injection from one layer to another is most effective within a certain range of iron layer thicknesses. 
\end{abstract}

%
%
%
%
\ioptwocol

\section{Introduction}

In the race to reduce power consumption and increase processing capability, the use of spin rather than charge promises a new generation of microelectronics. Spintronics based on metallic multilayers employs these structures for data storage and spin transport for information exchange. Conventional spintronics employs spin currents and spin densities that are due to non-equilibrium electrons that are still close to the Fermi energy.~\cite{Hirohata_2014}  

Even though metal-based spintronics is a very well established field, it continues to rapidly evolve. Among the current goals of spintronics research are the manipulation of spin ensembles carried by electrons that are energetically farther away from the Fermi energy.
For non-equilibrium electrons in general, scattering events determine the electronic transport in magnetic materials, regardless of the source that drives the currents, for instance, electrical spin injection from magnetic (metallic or semiconductor) electrodes \cite{Okamoto2016,Tombros2007,johnson2015}, or manipulating spin polarization by alternating-current (AC) magnetic fields via Zeeman interaction \cite{Zhang2018,Busl2010}, to name but a few. 
If one considers electron injection at high energies or, in particular, optical excitation of electronics, dynamics with an essential electronic energy dependence and its interplay with spin transport becomes more important. For this hot-electron spin transport an intermediate ``superdiffusive'' regime was identified~\cite{PhysRevLett.105.027203}, which has characteristics somewhat different from ballistic and diffusive transport. Based on experimental evidence~\cite{Rudolf2012, PhysRevLett.117.147203}, it is believed that superdiffusive spin currents can be launched from a ferromagnetic layer into adjacent metallic layers and can contribute to the transfer of spin angular momentum that is needed in the ultrafast demagnetization of ferromagnets. An alternative method to describe superdiffusive spin currents, is provided by a particle in-cell approach~\cite{PhysRevB.98.224416,PC_binder}. This method to solve the spin-dependent Boltzmann equation can be relatively easily adapted to ab-initio input.

In this work, we focus on the dynamics and transport of electrons created far away from the Fermi energy by fs laser irradiation. We study a prototypical Fe/Au bilayer system, where optical pulses, which drive the hot-electron spin currents, are absorbed in both the ferromagnetic and normal metal layers. Our approach is based on the Monte Carlo model for spin-dependent electron dynamics developed in Ref.~\cite{Briones_2022}. In addition to the effects of different collision processes, such as secondary electron generation and elastic scattering, we investigate the influence of different thicknesses of the magnetic layer and the interface transmissions with the intent to understand and optimize spin currents in the structure.

\section{Theoretical Approach}

The aim of our study is to predict how different interactions influence the nonequilibrium dynamics and spin injection in FM/M bilayer systems. In our approach, we consider free electron states above $E_\mathrm{F}$ as essentially free and focus on the influence of high-energy electrons in spin transport. This section presents the algorithms, equations and set of parameters that will be used for the simulations presented in this manuscript. 

\begin{figure}[H]
    
    \includegraphics[scale =0.7, center]{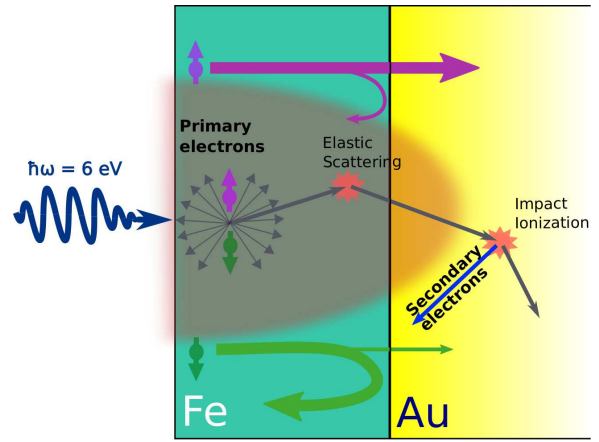}%

    \caption{Schematic representation of the laser irradiation and scattering events in the Fe/Au bilayer system. There are two possible interactions to consider: electron–atom interactions where only the direction of flight changes (elastic scattering) and electron-electron interaction generating secondary electrons (impact ionization). When electrons enter in contact with the interface, their spin determines their transmission probability, and they cross accordingly.}
    \label{fig:Scheme}

\end{figure}

\subsection{Monte Carlo method}\label{Sec::MC_method}

The asymptotic Monte Carlo trajectory method \cite{MC_asymp} is a statistical technique that models binary collision interactions by sampling several trajectories until an estimate of the possible outcomes is obtained. Probability theory is used to implement an algorithm for random sampling of variable $x$. In probability theory, all possible events $p(x)$ are integrated into a variable referred to as the cumulative distribution function (CDF) $F(x)$ in order to construct a formula that can be used to obtain a value for the variable $x$ \cite{MCparticle}. A more thorough and accurate description of how the Monte Carlo method is done can be found in literature \cite{alireza, Landau_MC,HUTHMACHER2015242}. Here, we will study the dynamics of excited particles using the same technique explained in our previous work \cite{Briones_2022}. During the simulation, any electron interaction process is treated by random sampling. The time between two successive collisions (time of free flight) $\tau$ can be sampled with the random variable $\mathrm{R} \in [0,1]$ as
\begin{equation}\label{eq:freetime}
 \tau = -\nu^{-1}_0\log (\mathrm{R}),
\end{equation}
when the scattering rate $\nu_0$ is constant. However, assuming a constant total scattering rate, independent of energy, is a statistical overestimation. To compensate, we introduce a further possibility with an energy-dependent probability, which allows the particle to continue its trajectory unperturbed. For the case of several scattering mechanisms as studied in this work, we perform a random sample of collisions using the probability function $p(x)$ which can be replaced with either the differential cross-section $d\sigma/d\Omega$, scattering rates $\nu$ or characteristic times $\tau$.

\subsection{Material parameters}\label{subsection::Laser_parameter}

We consider a Fe/Au bilayer material irradiated by a Gaussian-like laser pulse of $6\;\mathrm{eV}$ photon energy. The laser irradiates the material from the iron side, as shown in Fig. \ref{fig:Scheme} and the maximum laser pulse irradiation is centered at $\SI{0}{\femto\second}$. The iron layer thickness is finite, but the gold layer thickness, is taken to be infinite and the electrons are tracked only in the first $\SI{10}{\nano\meter}$. Excited electrons then interact within the material, either changing only their momentum (elastic scattering) and/or generating secondary electrons (impact ionization). As soon as an electron reaches the interface, its transmission to the other layer is determined by its spin and generally also its energy. 


There are two major parameters for this simulation that we tune and study: (i) The thickness of the Fe layer, 
and (ii) the transmission probability $T$ at the interface. We focus on the excitation of high-energy electrons where the transmission probability is rather insensitive to the electronic energy
(see figure 4 in \cite{PhysRevLett.119.017202}). Thus, we approximate the transmission probability as a constant.  
and also assume it to be independent of propagation direction of the electrons reaching the interface
i.e., it is assumed to be identical for
electrons traveling from iron to gold and the other way around. The constant transmission probability  will be taken taken from Ref. \cite{PhysRevLett.119.017202} as $T^{\uparrow} = 0.95$, and $T^{\downarrow} = 0.25$ for spin-up and spin-down electrons, respectively. 



    
    
    

\section{Scattering processes}

We trace the dynamics of optically excited electrons by modeling pure jump processes~\cite{FoundationMC, SimulationMC,HUTHMACHER2015242}. We consider two kinds of interactions in this study, namely the electron-atom interaction, which is treated as an elastic scattering process here, and the electron-electron impact ionization, in which energy is transferred to a secondary electron through an inelastic process. 

Elastic scattering processes will  be modeled as a deflection  without transfer of energy. As indicated in subsection \ref{Sec::MC_method}, the probability function $p(x)$ will be replaced by the differential cross-section based on Mott's cross-section in dependence of the scattering angle. We will use $\tau_{el} = 25\;\mathrm{fs}$ as the elastic scattering time~\cite{PhysRevB.71.233104}, which was applied to study spin transport and spin dynamics in Refs.~\cite{PhysRevB.98.224416,PhysRevB.85.235101}.

Electron impact ionization is a process in which incident electrons ionize electrons from occupied states into unoccupied states. The newly generated high-energy electrons are referred to as secondary electrons, whereas the optically excited electrons are referred to as primary electrons. With the energy lost by the primary electron of $\Delta E$  and the binding energy $I$ of the bound electron, the final energy of the newly ionized electron $E_s$ above Fermi level is $E_s = \Delta E - I$. The amount of transferred energy is assumed to be half of the energy of the incident electron as it was done in Ref.~\cite{RITCHIE19651689}. For the inelastic scattering rate ($\tau_{ee}^{-1}$) we will consider the energy-dependent collision rate of an excited electron at temperature $T_{e}$~\cite{PhysRevB.87.035139}. Both collision processes and their inclusion are described in more detail in Ref. \cite{Briones_2022}.

\section{Results}

The behavior of the nonequilibrium spin transport and spin dynamics will be analyzed in a  Fe/Au bilayer system after fs-laser excitation. The laser pulse irradiates the sample from the iron layer (see Fig. \ref{fig:Scheme}), and the distance to the interface is measured from this point inwards. Therefore, the thickness of the iron layer will determine the location of the boundary between these materials. Throughout this work, we will use a variety of iron thicknesses, and they will be specified when required. As for the thickness of the gold layer, it is always infinity and we trace the dynamics within the first 
$\SI{10}{\nano\meter}$. 
The effects on the spin transport can be addressed by studying the spin currents generated at the interface of the material. The spin current density $j_s$ is defined as
\begin{equation}\label{eq::Spcrr}
    j_s(z,t)\propto q[ \langle \eta^{\uparrow} v_\uparrow \rangle - \langle \eta^{\downarrow} v_\downarrow \rangle ] 
\end{equation}
where $q$ is the charge of the electron, $\eta^{\uparrow}$ ($\eta^{\downarrow}$) and $v_\uparrow$ ($v_\downarrow$) are the particle density and the velocity for spin up (spin down), respectively. An average over the number of electrons at a specific volume and time is denoted by the angle brackets. Both expressions in eq. \eqref{eq::Spcrr} can be separated as:
    
\begin{eqnarray}
    J_{up} &\propto  &\langle \eta^{\uparrow} v_\uparrow \rangle \label{eq:jup}\,,\\
    J_{down} &\propto  &\langle \eta^{\downarrow}  v_\downarrow \rangle\,\label{eq:jdown}.        
\end{eqnarray}

\subsection{Spin current density in bulk iron}

We start by examining the behavior of the spin current density in bulk iron. We then extend the analysis to a more complex system, namely the Fe/Au bilayer. As it was analyzed in Ref.~\cite{Briones_2022}, particles reach an average kinetic energy of less than $\SI{2}{\eV}$ very rapidly because secondary electrons are 
likely to be generated during the initial time period of the nonequilibrium dynamics. 


\begin{figure}[H]    
\centering
    \includegraphics[scale =0.68, center]{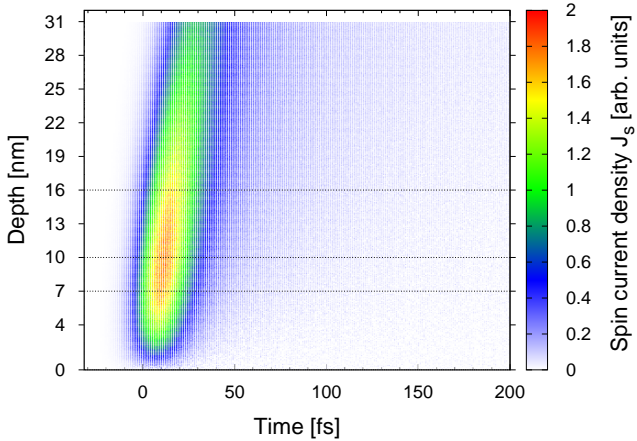}
    
    \caption{Development of the spin current density $j_s$, defined in eq. \eqref{eq::Spcrr}, in bulk iron over time and in depth, considering secondary electron generation. The dashed lines indicate the position where the spin current density is evaluated further.}
    \label{fig:Spin_current}
    
\end{figure}

First, let us focus on the irradiation of bulk iron by femtosecond laser pulses as it was described in reference \cite{Briones_2022}. A simulation is performed using bulk-like Fe film which is excited by a Gaussian-like fs laser pulse with a photon energy of $\SI{6}{\eV}$. 

As a result of the laser excitation, majority and minority electrons are brought out of equilibrium with different kinetic energy distributions, and also random initial directions. We use open boundary conditions at the end of the calculation region of $\SI{30}{\nano\meter}$ thickness, allowing electrons to escape completely from the simulated system.

Figure \ref{fig:Spin_current} shows the evolution of the spin current density, as defined by eq.~\eqref{eq::Spcrr}, as a function of time and depth. Note that the magnitude of spin current density remains positive over time and throughout the material. This indicates that, on average, majority electrons move towards the depth of the material. In addition, the magnitude of spin current density increases steeply with increasing depth and reaches a maximum value at a around $\SI{10}{\nano\meter}$, which will be analyzed later.

\begin{figure}[H]    
\centering
    \includegraphics[scale =0.7, center]{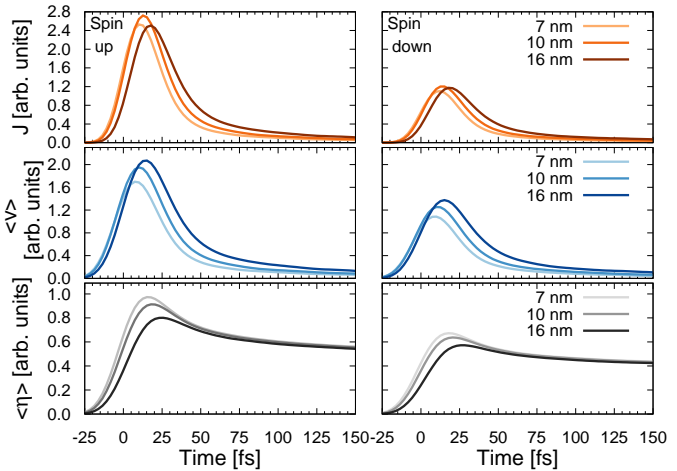}
    
    \caption{An analysis of the evolution of the spin up (lhs) and spin down (rhs) current density $J$ (top graph), average velocity $v$, and particle density $\eta$ in bulk iron, as defined in eq. \eqref{eq::Spcrr}. The physical quantities are depicted for different depths at the location of the dashed lines in fig. \ref{fig:Spin_current}.}
    \label{fig:etavel_J_BL}
    
\end{figure}

A more detailed observation of the trend in the spin current density is provided in fig. \ref{fig:etavel_J_BL}. Here, we depict the contributions to the spin current density in time at different depths, as indicated in fig. \ref{fig:Spin_current}. The top graph represents the spin current density for both spin up (lhs) and spin down (rhs) electrons, the middle graph depicts the average velocity $\langle v \rangle$, and the bottom graph displays the average particle density $\langle \eta \rangle$, terms used in eqs. \eqref{eq:jup} and \eqref{eq:jdown}. The color gradient indicates at what depth the variables were selected for analysis. The variables shown in fig. \ref{fig:etavel_J_BL} all seem to exhibit the same behavior in time, but a different maximum value for different depths. We observe again how the average particle density $\langle \eta \rangle$ shows a homogeneous density distribution further in time, which was already discussed in ref. \cite{Briones_2022}. 


\begin{figure}[H]    
\centering
    \includegraphics[scale =0.68, center]{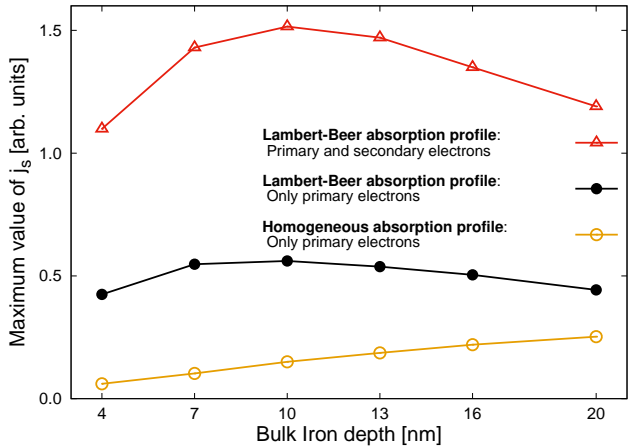}
    
    \caption{Maximum values of the spin current density $j_s$ in bulk iron at different depths, for different absorption profiles: Homogeneous absorption profile exciting only primary electrons (yellow curve), Lambert-Beer absorption profile with only primary excitations (black curve), and including primary and secondary electrons (red curve). The values for $j_s$ were obtained using eq. \eqref{eq::Spcrr}.}
    \label{fig:MaxJ_iron}
    
\end{figure}

Based on these simulations, we have identified two possible causes for this behavior, either the type of scattering or the absorption profile. In fig. \ref{fig:MaxJ_iron}, it is shown how the maximum values of spin current density $j_s$ are affected by the type of scattering and absorption profile. The red curve has been calculated using Beer-Lambert's absorption probability to excite primary electrons and including secondary electron generation. The black curve was done by keeping the same absorption profile, but considering only optically excited electrons. The magnitude of the maximum spin current density $j_s$ in both graphs increases with depth. This increment then slows down until a certain depth (around $\SI{10}{\nano\meter}$), after which it gradually decreases. These two graphs then differ only qualitatively, which is in agreement with ref. \cite{Briones_2022}, where it is demonstrated that secondary electrons affect spin current density quantitatively rather than qualitatively. As we consider that the absorption profile influences the observed behavior at different maxima of spin current density, we need to examine the effect of a different profile on primary electrons. This particular scenario involves a homogeneous absorption profile, i.e., there will be the same probability of absorption at every depth. 

As shown in fig. \ref{fig:MaxJ_iron}, contrary to other curves, where the maximum values increase until a certain depth, and then gradually decrease (red and black curves), the maximum value of the spin current density $j_s$ in a non-interactive picture (yellow curve), where only elastic scatterings are taking place, increases steadily in depth. This trend can be explained by examining the dynamics of particles following laser absorption. In the non-interactive picture (yellow curve), particles are initially excited based on a constant absorption profile throughout the material. However, as they move towards the depth of the material, the distribution of particles changes since their directionality is initially random. It is likely that particles will move ballistically in the absence of further interactions (change in direction or generation of new particles), and that the density of particles within the material will not be significant. Due to their initial distribution when they are created as well as their interactions with each other, particles, over time, acquire a common directionality through interaction. Hence, the absorption profile is responsible for the observed tendency in the maximum spin current density values in bulk iron at different depths.

    
    


\subsection{Spin current density in Fe/Au bilayer system}

\begin{figure}[H]  
  \includegraphics[scale =0.6, center]{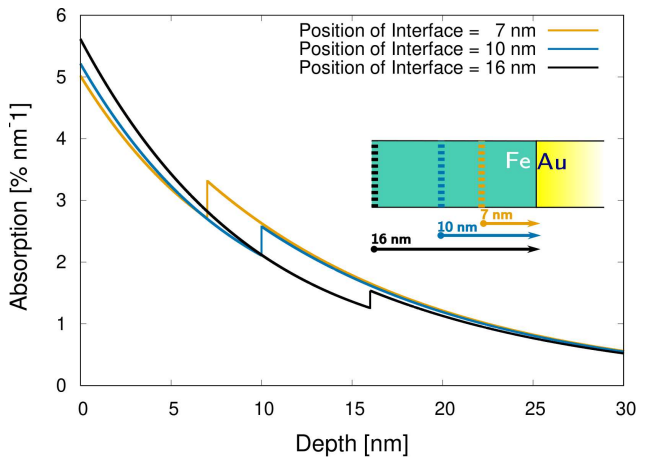}
   
  \caption{Absorption probability with respect to the depth for Fe/Au bilayer system with an interface at $\SI{7}{\nm}$ (yellow), $\SI{10}{\nm}$ (blue) and $\SI{16}{\nm}$ (black). Both materials have a different absorption profile. At the interface, one can observe a sudden increase in the absorption when the energy enters into the gold layer.}
   
\label{fig:Prob_depth}
\end{figure}

We now proceed to examine the spin transport in a bilayer system containing iron and gold. Unlike bulk iron, laser radiation induces primary excitations within both layers in accordance with the absorption profile and these are initially distributed within the material accordingly, as illustrated in fig. \ref{fig:Prob_depth}. Fig. \ref{fig:Prob_depth} shows the data of the laser light absorption probability for different thicknesses of iron layers, which  determine the position of the interface in the bilayer system: $\SI{7}{\nano\meter}$ (yellow curve), $\SI{10}{\nano\meter}$ (blue curve) and $\SI{16}{\nano\meter}$ (black curve). These curves were obtained following the procedure described in ref.\cite{Eschenlohr,Khorsand2014} with absorption coefficients extracted from ref.\cite{Wolfgang2009}. The reason behind choosing these specific sizes is to study how smaller or larger iron layers
influence the dynamics. The data in fig. \ref{fig:Prob_depth} was calculated for a semi-infinite system, to avoid the reflective effects from gold.

Over time, primary electrons generate secondary electrons, and these subsequently generate further electrons, a process that is repeated in both materials. At the boundary between iron and gold, which will be later addressed as interface, particles are transmitted in different proportion due to the spin dependent transmission probability (spin filtering effects). 

\begin{figure}[H]    
\centering
    \includegraphics[scale =0.7, center]{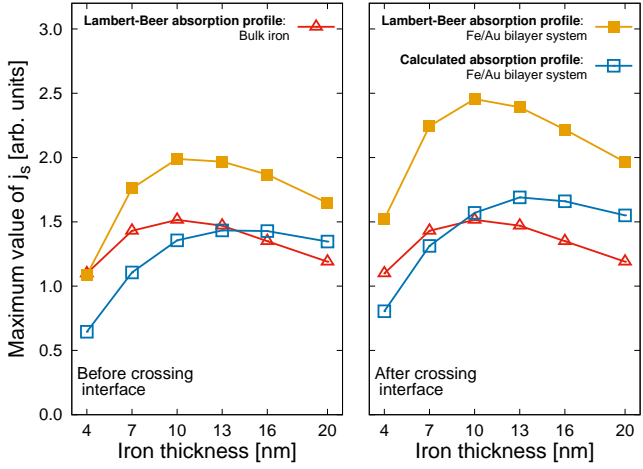}
    
    \caption{A comparison of the maximum values of the spin current density $j_s$, obtained from eq. \eqref{eq::Spcrr}, for different thicknesses of iron. The value is compared between the Lambert-Beer absorption profile in bulk iron (red) and in the Fe/Au bilayer system (yellow), as well as one simulation using a calculated absorption profile (blue), applying eqs. from refs. \cite{Khorsand2014,Eschenlohr} and parameters in ref. \cite{Wolfgang2009}.}
    \label{fig:Max_J}
    
\end{figure}

In light of our previous analysis of iron, we now have a better understanding of how the absorption profile influences the behavior of the spin current density. Fig. \ref{fig:Max_J} illustrates how the absorption profile and thickness of iron layers affect the maximum value of spin current density in a bilayer system before and after crossing the interface. The magnitude of the peak of the spin current density is compared between the Lambert-Beer absorption profile in bulk iron (red curve), in the Fe/Au bilayer system (yellow curve) and using a calculated absorption profile (blue curve) (using parameters, procedure and absorption profile from refs. \cite{Wolfgang2009,Eschenlohr,Khorsand2014}, respectively).

\begin{figure}[H] 
    \centering
    \includegraphics[scale =0.66, center]{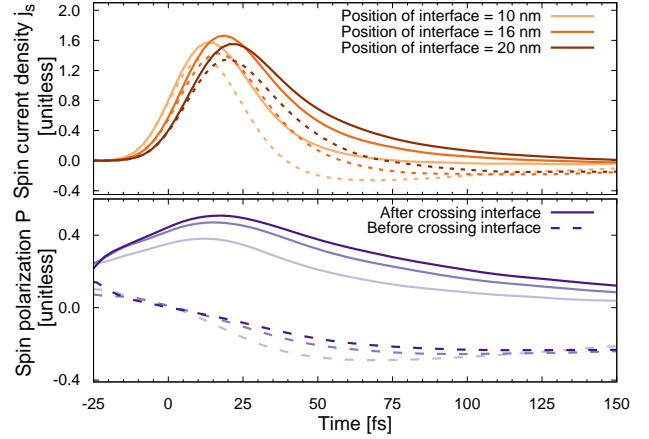}
       
    \caption{Spin current density $j_s$ (top graph) and spin polarization $P$ (bottom graph), calculated before (dashed lines) and after (solid lines) crossing the interface. The different thicknesses of the iron layers are displayed as color degraded lines.}
        
    \label{fig:SpCrr_pol}
       
\end{figure}

In fig.~\ref{fig:Max_J} Iron (red) and Fe/Au (yellow) exhibit a large difference in maximum spin-current density as a result of the interface, which increases the value of $j_s$. However, the interface does not alter the tendency observed over a preferential thickness of iron layer. However, when the calculated absorption profile (from fig. \ref{fig:Prob_depth}) is taken, the preferred thickness changes and the larger spin current injection from iron to gold appears at a lower iron thickness. By analyzing the Lambert-Beer (yellow) and the calculated (blue) absorption profiles, one realizes the preferable thickness changes. This means that because of the influence of light absorption in the bilayer system, there is a specific thickness at which a larger spin current injection from iron to gold is observed. Spin dependent particles that cross between layers reconfigures the distribution of particles in gold, which influences the spin current density. This can be analyzed using the equation for spin polarization $P$ : 
\begin{equation}\label{eq:spin_pol}
    P = \frac{\eta^{\uparrow} - \eta^{\downarrow}}{\eta^{\uparrow} + \eta^{\downarrow}},
\end{equation}
where $\eta^{\uparrow}$ ($\eta^{\downarrow}$) is the particle density of spin up (down) particles. Fig. \ref{fig:SpCrr_pol} shows the spin current density $j_s$ (top graph) and the spin polarization $P$ (lower graph), calculated using eqs. \eqref{eq::Spcrr} and \eqref{eq:spin_pol}, respectively, for three different thicknesses of iron layers.

The dashed lines indicate quantities taken before crossing the interface, while the solid lines indicate quantities taken after crossing the interface. The lines depicting the different thicknesses of the iron layer are color degraded. For the spin current density, the observed behavior is similar to that of fig.~\ref{fig:etavel_J_BL}. In this bilayer system, we focus on the thickness of the iron layer instead of the propagation depth. As shown in fig.~\ref{fig:SpCrr_pol}, the maximum value of the spin current density $j_s$ increases up to a certain thickness (here $\SI{16}{\nano\meter}$), but decreases for a thicker film. As a result, it appears that the maximum value of spin current density does not increase with increasing iron layer thickness. Instead, there exists an optimal size of iron layer at which spin injection from iron to gold is more effective. The violet curves in fig.~\ref{fig:SpCrr_pol} are the spin polarization curves, which were calculated using eq. \eqref{eq:spin_pol}. They indicate that in time, always more particles with spin up will cross the interface as a result of the difference in transmission probability for the spins. By increasing the size of the iron layer, the maximum value of spin polarization will be reached at later times. This delay is a consequence of the time necessary for particles to interact within the material and travel toward the interface. As the thickness of iron layer increases, spin polarization increases since fewer particles cross over to gold when the layer is larger.

\section{Summary}

In conclusion, we investigated how different iron thicknesses affect the generation of spin currents at the interface of a Fe/Au bilayer system.  We first examined the effects of the light absorption profile on the spin current density within a bulk iron layer, and we found that the absorption profile influences the spin current density non monotonously. 
For the Fe/Au bilayer, we computed the nonequilibrium transport of excited electrons and studied how different absorption profiles influence the spin current injection from one layer into another. 
This analysis also revealed that secondary electron generation and the interface affect spin transport only in terms of magnitude. We then investigated the combined effect of light absorption and spin-dependent interface transmission probability  on the spin current density in the Fe/Au bilayer system. 
The magnitude of the spin current density from Fe to Au is influenced by the thickness of the Fe layer and shows a maximum at intermediate thicknesses. We found that the absorption characteristics of the exciting laser pulse has a direct influence on the
efficiciency of the spin injection from one layer into another.

\ack Funded by the Deutsche Forschungsgemeinschaft (DFG, German Research Foundation) - TRR 173/2 - 268565370 Spin+X (Project B03). B.S. acknowledges financial support by the Dynamics and Topology Center, funded by the State of Rhineland-Palatinate.

\section*{References}

\bibliographystyle{iopart-num}
\bibliography{biblio}

\end{document}